\documentclass[aps,prb,showpacs,superscriptaddress,twocolumn,longbibliography]{revtex4-2}

\usepackage{amsmath}
\usepackage{graphics}
\usepackage[next]{inputenc}
\usepackage[dvips]{epsfig}
\usepackage[colorlinks=true,citecolor=blue,linkcolor=blue]{hyperref}
\usepackage{bbm}
\usepackage{bbold}
\usepackage{booktabs}
\usepackage{multirow}
\usepackage{hhline}
\usepackage{graphicx}
\usepackage{amsmath}
\usepackage{multirow}
\usepackage[usenames, dvipsnames]{color}
\usepackage{color,soul}

\def\be{\begin{equation}}
	\def\ee{\end{equation}}

\def\bi{\begin{itemize}}
	\def\ei{\end{itemize}}
\def\bn{\begin{enumerate}}
	\def\en{\end{enumerate}}
\def\bea{\begin{eqnarray}}
	\def\eea{\end{eqnarray}}

\def\ba{\begin{array}}
	\def\ea{\end{array}}
\def\bd{\begin{displaymath}}
	\def\ed{\end{displaymath}}

\begin{document}
	
	\title{Diverse Magnetic Phase Diagram and Anomalous Hall Effect in Antiferromagetic LuMn$_6$Sn$_6$}
	
	\author{Shirin Mozaffari}
	\email{smozaff1@utk.edu}
	\affiliation{Department of Materials Sciences and Engineering, The University of Tennessee, Knoxville, TN 37996, USA}

   \author{Seung-Hwan Do}
      \affiliation{Department of Materials Sciences and Engineering, The University of Tennessee, Knoxville, TN 37996, USA}
 
	\author{Richa P. Madhogaria}
	\affiliation{Department of Materials Sciences and Engineering, The University of Tennessee, Knoxville, TN 37996, USA}

	\author{Aikaterini Flessa Savvidou}
	\affiliation{National High Magnetic Field Laboratory, Florida State University, Tallahassee, Florida 32310, USA}

	\author{Brian W. Casas}
	\affiliation{National High Magnetic Field Laboratory, Florida State University, Tallahassee, Florida 32310, USA}
 
        \author{William R. Meier}
	\affiliation{Department of Materials Sciences and Engineering, The University of Tennessee, Knoxville, TN 37996, USA}
  
	\author{Rui Xue}
	\affiliation{Department of Physics and Astronomy, The University of Tennessee, Knoxville, TN 37996, USA}
	
	\author{Eun Sang Choi}
	\affiliation{National High Magnetic Field Laboratory, Florida State University, Tallahassee, Florida 32310, USA}
	
	\author{Luis Balicas}
	\affiliation{National High Magnetic Field Laboratory, Florida State University, Tallahassee, Florida 32310, USA}
	
	\author{David G. Mandrus}
	\email{dmandrus@utk.edu}
	\affiliation{Department of Materials Sciences and Engineering, The University of Tennessee, Knoxville, TN 37996, USA}
	\affiliation{Department of Physics and Astronomy, The University of Tennessee, Knoxville, TN 37996, USA}

	\begin{abstract}
        The interactions between conduction electrons and magnetism can significantly enhance the Hall signal, a phenomenon known as the anomalous Hall effect (AHE). While the AHE is generally not expected in antiferromagnets, a large AHE is observed in certain antiferromagnets with noncollinear spin textures and nonvanishing Berry curvature. In this work, we present a rich temperature and magnetic phase diagram with eight distinct magnetic phases for the antiferromagnetic kagome compound LuMn$_6$Sn$_6$. The Hall effect analysis in LuMn$_6$Sn$_6$ reveals both intriguing physical phenomena and methodological challenges.         
        In the coplanar canted antiferromagnetic phase, we observe an AHE, which likely originates from the intrinsic effects. 
        At low temperatures, upon entering the ferromagnetic phase, the AHE sharply increases and exceeds the conventional limits expected from intrinsic mechanisms.
        We also demonstrate the limitations of standard experimental methods in extracting the topological contribution to the Hall effect data. We show the importance of considering magnetoresistance anisotropy when estimating the anomalous and topological Hall effects. 
        These shortcomings in current approaches in partitioning the Hall response necessitate new tools to interpret transport results in complex magnetic materials such as LuMn$_6$Sn$_6$.
       

                      

                
		
	\end{abstract}
	\date{\today}
	
	\maketitle
	\section{Introduction \label{intro}}
    The electrical Hall effect serves as a widely used probe to explore topological effects in magnetic systems. The Hall current requires breaking time-reversal symmetry. When this symmetry is broken by an external magnetic field, it results in the ordinary Hall effect (OHE).  Alternatively, the magnetic field can induce internal magnetization, resulting in an additional contribution to the Hall current, referred to as the anomalous Hall effect (AHE)~\cite{AHE_Allan,AHE_AFM}.      
    Intrinsic contributions to the AHE originate from the Berry curvature of the Bloch wave function, summed over all occupied states, analogous to the mechanism of the quantum Hall effect. 
   
    In ferromagnets with spontaneous magnetization, AHE can arise without the presence of an external magnetic field. Traditionally, the Hall effect has not been associated with antiferromagnetic (AF) order. However, recent theoretical and experimental studies have revealed significant Hall effects in certain magnetic crystals lacking global magnetization. Examples include the observation of a Hall effect in a spin-liquid candidate Pr$_2$Ir$_2$O$_7$~\cite{AHE_Pr2Ir2O7}, certain non-collinear AF compounds such as Mn$_3$Sn, Mn$_3$Ge, MnGe, NbMnP, and Ce$_2$CuGe$_6$~\cite{AHE_Mn3Sn,AHE_Mn3Ge,AHE_MnGe,AHE_NbMnP,AHE_Ce2CuGe6}, and the newly discovered Hall effect in altermagnetic materials~\cite{altermagnet_PRX}. Remarkably, the AHE in some of these antiferromagnets is comparable in strength to that observed in ferromagnets.

    These discoveries led us to move away from viewing the Hall effect solely as arising from conventional symmetry-breaking processes. Instead, the Hall response is now understood to be closely tied to non-trivial topology in the electronic structure, resulting in an enhanced response that challenges the traditional paradigm of scaling with total magnetization strength~\cite{AHE_AFM}. 

    In addition to the ordinary and anomalous Hall effects, noncoplanar spin alignments in real space can give rise to a finite scalar spin chirality, $\chi_{ijk} \sim s_i \cdot (s_j \times s_k)$, where $s_{i,j,k}$ are the spin moments of neighboring sites. This spin chirality can act as a local emergent magnetic field for conduction electrons, leading to a transverse response, termed the topological Hall effect (THE)~\cite{THE_Taillefumier,THE_MnSi,THE_chirality,THE_MnGe,Gd2PdSi3_THE}. 

	Among the numerous compounds with magnetic topological behavior, the family of kagome structure materials is a quintessential embodiment of the concept. Examples include Fe$_3$Sn$_2$~\cite{AHE_Fe3Sn2,AHE_Fe3Sn2_2}, Fe$_3$Sn~\cite{AHE_Fe3Sn}, Mn$_3X$ ($X=$Sn,~Ge,~Pt,~Rh,~Ir)~\cite{AHE_Mn3Sn,AHE_Mn3Sn2_2,Yang_AHE}, Co$_3$Sn$_2$S$_2$~\cite{AHE_Fe3Sn2_Co3Sn2S2}, and the family of $R$Mn$_6$Sn$_6$ ($R$=~Rare earth)~\cite{R166_progress}. The latter family, which is strongly frustrated, is the focus of this paper. 

    \begin{figure}[hbt]
		\includegraphics[width=\linewidth]{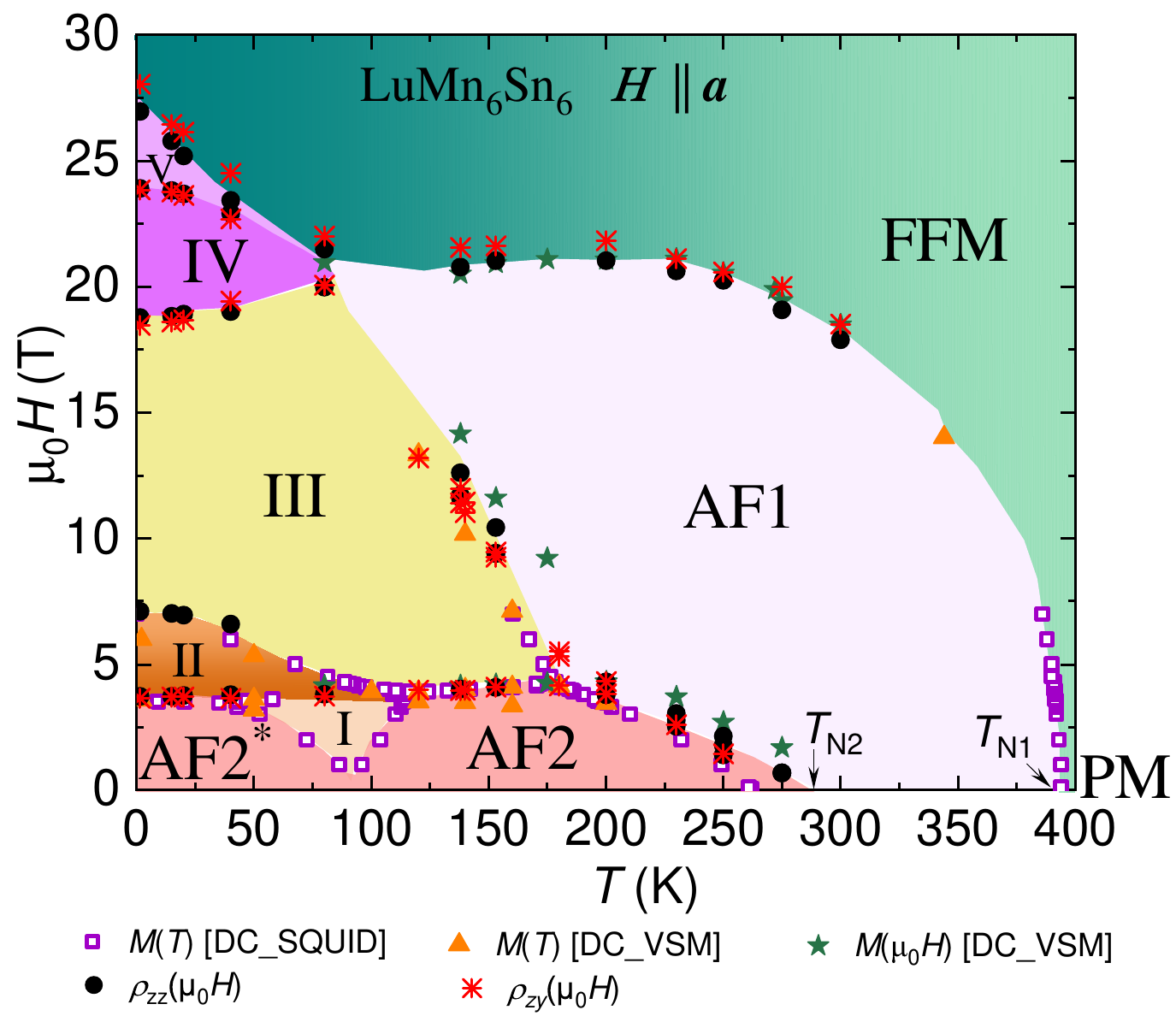}
		\caption{Temperature ($T$) and magnetic field ($\mu_0H$) phase diagram of LuMn$_6$Sn$_6$ for field applied along the crystallographic \textit{a}-axis. A variety of field-driven magnetic states are observed, including a collinear antiferromagnetic (AF1) and a forced ferromagnetic (FFM) phases. Under applied field, AF1 adopts a canted antiferromagnetic structure. Both AF2 and AF2$^*$ have a double spiral (DS) incommensurate structure, but with different periodicity~\cite{Hwan}. 
        Phases I and II show a commensurate antiferromagnetic order; Phase III adopts a transverse conical spiral (TCS) structure~\cite{Hwan}. Comparison with the phase diagram of YMn$_6$Sn$_6$~\cite{YMn6Sn6_Rebecca,YMn6Sn6_Lifshitz} suggests that phase IV corresponds to a fan-like (FL) structure. The ($H$, $T$) phase diagram was constructed using DC magnetization ($M$), longitudinal ($\rho_{zz}$) and Hall resistivity ($\rho_{zy}$) measurements.  
        } \label{phsses}
	\end{figure}

    $R$Mn$_6$Sn$_6$ family materials consists of kagome bilayers of Mn atoms that are separated by Sn and $R$ atoms along the \textit{c} axis. The family shows a collinear ferrimagnetic ground state with magnetic rare earths $R$~=~Gd--Er due to the strong AF coupling between $R$ and Mn atoms~\cite{RMn6Sn6_Liqin,TbMn6Sn6_Nirmal,Er_yazdi,Er_fazel,GdMn6Sn6,Dy_Ho,Ho,RMn6Sn6_highField}. 
    The non-magnetic rare earth analogues, $R$~=~Sc,~Y,~Lu, are antiferromagnets and consist of ferromagnetic (FM) double layers of the Mn kagome plane, which antiferromagnetically couple with neighboring bilayers~\cite{neutron1,neutron2,LuMn6Sn6_LuMn6Ge6,magnetization}. 
    
    $R$Mn$_6$Sn$_6$ is reported to have several field-induced magnetic phases, and the observation of THE in one of these phases in $R$=Y, Sc, Er is among the most exciting phenomena reported for this family~\cite{Nirmal,YMn6Sn6_Pengcheng,ScMn6Sn6_APL,Er_fazel}. 
    Beyond THE, $R$Mn$_6$Sn$_6$ compounds have demonstrated other intriguing phenomena, including magnetization-driven Lifshitz transitions~\cite{YMn6Sn6_Lifshitz}, a large anomalous transverse thermoelectric effect~\cite{YMn6Sn6_thermoelectric}, and topological Nernst and thermal Hall effects~\cite{Richa2023,YMn6Sn4Ge2_Nernst}.
    Among the members of $R$Mn$_6$Sn$_6$, LuMn$_6$Sn$_6$ is studied the least likely because of the very strong field required to induce the fully polarized state.

    In this study, we delineate a rich field-temperature phase diagram for LuMn$_6$Sn$_6$ under a magnetic field applied along the crystalline \textit{a}-axis, as shown in Fig.~\ref{phsses}.
    Such a rich phase diagram is rare and indicates the competition between multiple interactions. We identified eight different magnetic phases as a result of competition between FM and AF coupling between Mn kagome layers. To label the numerous magnetic phases, Roman numerals are used throughout this paper.

    In addition to the diverse magnetic phase diagram, our Hall effect measurements shows an enhancement of an AHE contribution across the AF2 -- AF1 transition and at the boundary between phase IV (V) and the FFM state. Moreover, our study advocates for theoretical models over empirical relations and underscores the importance of considering magnetoresistance anisotropy in the analysis of nontrivial Hall responses. LuMn$_6$Sn$_6$ is ordered magnetically at room temperature and thus potentially has a direct application in the design of complex magnets for data storage.

    \section{Experimental Details} \label{Exp}
	Crystals of LuMn$_6$Sn$_6$ were grown from a tin flux using an atomic ratio of Lu:Mn:Sn = 1:6:30. The Lu pieces (Alfa Aesar 99.9\%), Mn pieces (Alfa Aesar 99.95\%) and the Sn shot (Alfa Aesar 99.9999\%) were loaded into an alumina crucible. 
    The crucible assembly was sealed in a fused silica ampule and heated to 973~$^\circ$C over 12 h, then held for 12~h. The mixture was then cooled to 923~$^\circ$C at a rate of 2.5~$^\circ$C/h, followed by heating to 963~$^\circ$C over 2~h. Finally, it was cooled to 600~$^\circ$C at a rate of 1.5~$^\circ$C/h. At this temperature, the flux was separated from the crystals by inverting the tube and centrifuging.  This process yielded blocky hexagonal crystals up to 1~cm in size [Fig.~\ref{crystal}(a)]. The additional heating step to 963~$^\circ$C was included to promote the growth of larger crystals.     
    
    Powder X-Ray Diffraction measurement was performed using a Cu \textit{K}$_\alpha$ X-ray source.
    Rietveld refinements confirms the \textit{P6/mmm} HfFe$_6$Ge$_6$-type structure of the sample (see Fig.~S1 in the Supplemental Material (SM)~\cite{supp}). 
    Conventional magnetotransport experiments were performed in a physical property measurement system (PPMS-Quantum Design) under magnetic fields up to 14~T and temperatures as low as 1.8~K. Magnetization measurements under fields up to 7~T were performed in a commercial superconducting quantum interference device magnetometer (SQUID-Quantum Design). 
    The high field magnetization and magnetotransport measurements were performed in a resistive Bitter magnet at the National High Magnetic Field Laboratory (NHMFL) in Tallahassee, FL, under continuous fields up to 35 T. A vibrating sample magnetometer (VSM) was used to measure DC magnetization. 
    The measured longitudinal and transverse resistivities were field-symmetrized and antisymmetrized, respectively, to correct the effect of contact misalignment. 
	
	\section{Results and Discussion}
    \subsection{Crystal Structure Analysis\label{XRD}}
	\begin{figure*}[hbt]
		\includegraphics[width=1\linewidth,trim={0 0cm 0 0}]{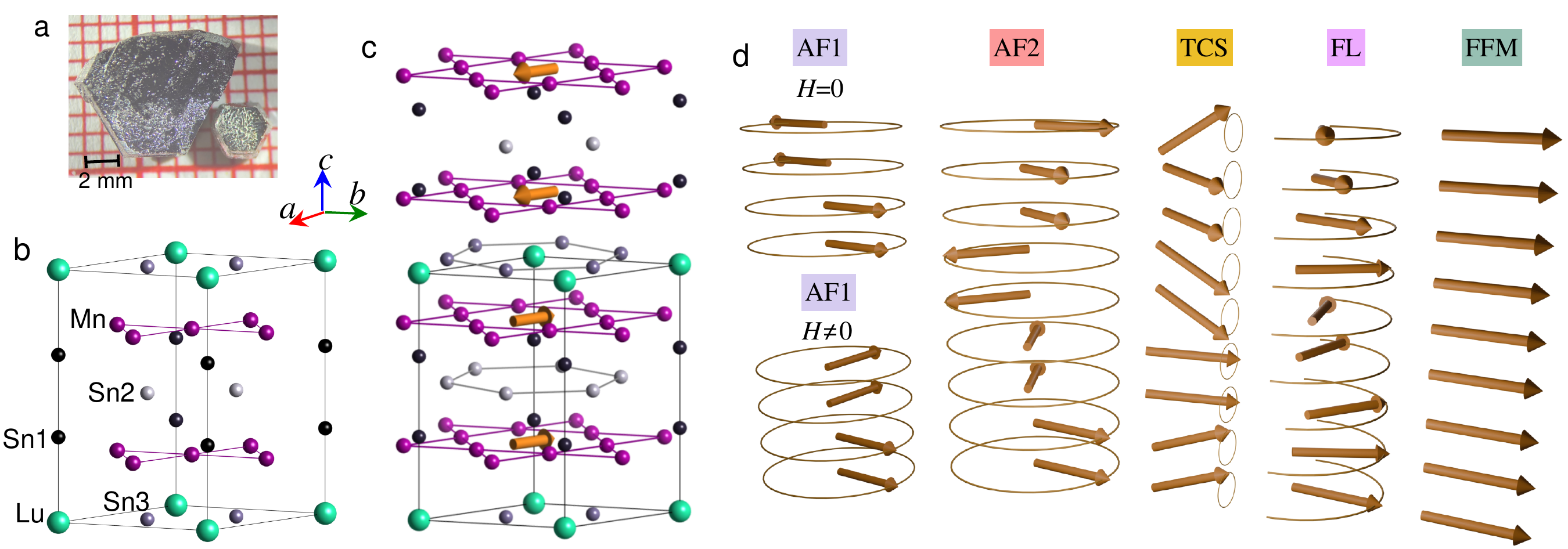}
       \caption{(a) A picture of typical hexagonal prismatic LuMn$_6$Sn$_6$ crystals. (b) Sketch of LuMn$_6$Sn$_6$ structure generated with Vesta. The boxed area shows the primitive cell of the crystal structure. The three Sn positions are shown with three different shades of grey color. (c) Room temperature  magnetic structures (AF1 phase) of LuMn$_6$Sn$_6$ at zero field  based on the neutron diffraction measurements~\cite{neutron2}. For simplicity, the magnetization of each manganese kagome sheet is shown by an arrow in the middle. (d) Sketch of different magnetic structures for YMn$_6$Sn$_6$ induced by a field applied along the \textit{a} crystalline axis~\cite{Nirmal,YMn6Sn6_Rebecca}. Arrows reoresent orientation of Mn moments within kagome sheets.}  \label{crystal}
	\end{figure*} 
	
	LuMn$_6$Sn$_6$ has a hexagonal structure with space group $P6/mmm$ (No. 191), as shown in Fig.~\ref{crystal}(b). The structure can be viewed as stacking layers of Lu, Mn, and Sn. Within each unit cell there are three different Sn sites (Sn1, Sn2, and Sn3) and two 
    kagome planes of Mn atoms which are separated by different layers of LuSn3 and Sn1-Sn2-Sn1. Only Mn atoms possess magnetic moments which are aligned ferromagnetically within the \textit{ab}-planes of the hexagonal lattice; the average of which in each Mn layer is shown schematically with an orange arrow in Fig.~\ref{crystal}(c). Neutron diffraction in zero external magnetic field shows that the Mn layers across the Sn1-Sn2-Sn1 layers are coupled ferromagnetically but the coupling through the Lu-Sn3 layer is antiferromagnetic~\cite{neutron1,neutron2}, as depicted in Fig.~\ref{crystal}(c). This makes the magnetic unit cell in the AF1 phase twice the size of the chemical unit cell. 
    The general magnetic structure in the $R$Mn$_6$Sn$_6$ ($R$ = Sc, Lu, Y) family and the band topology depend on the $R$ element type. Table~\ref{table} summarizes the lattice parameters and N\'eel temperatures in this family. The \textit{a} and \textit{c} lattice parameters of the Lu compound are in between the Sc and Y compound. 
	\begin{figure*}[hbt]
		\includegraphics[width=\linewidth]{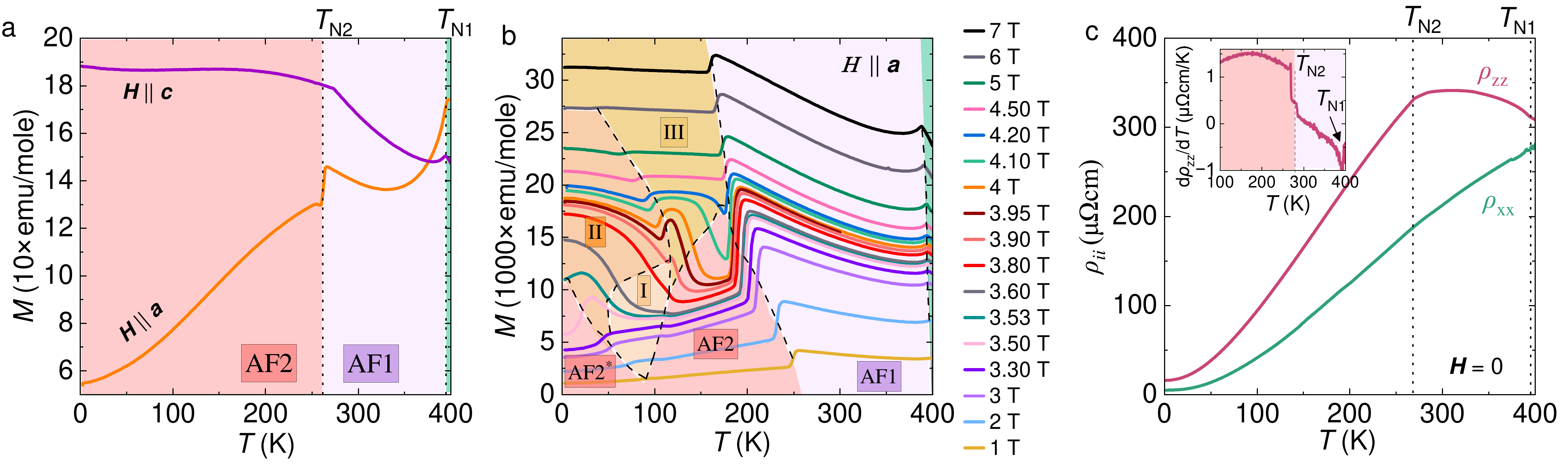}
		\caption{(a) Magnetization \textit{M} as a functions of temperature \textit{T} for magnetic field $\mathrm{\mu_0} H=$ 0.05 T applied parallel to \textit{a} and \textit{c}- axis of the LuMn$_6$Sn$_6$ crystal, obtained on heating the sample after zero-field-cooling (ZFC) down to $\sim 2$ K. (b) $M(T)$ curves measured under several different magnetic field values for $\mathrm{\mu_0} H$ applied along a axis. Dashed lines define the boundaries between different field-induced magnetic phases. (c) Electrical resistivity as a function of temperature, with the electric current applied along the \textit{a}-axis ($\rho_{xx}$) and \textit{c}-axis ($\rho_{zz}$) of the crystal. Inset shows the temperature derivative of $\rho_{zz}$ in the vicinity of $T_\mathrm{N1}$ and $T_\mathrm{N2}$.  Dotted lines in (a) and (c) show antiferromagnetic transitions happening at $T_{\textrm{N1}}$ and $T_{\textrm{N2}}$. } \label{M-R}
	\end{figure*} 
	
        \begin{table}[]
            \caption{\label{table} Lattice parameters and the N\'eel temperatures for $R$Mn$_6$Sn$_6$, $R$~=~{Sc, Y, Lu}. $T_{\textrm{N1}}$ marks the transition to a collinear antiferromagnetic state and $T_{\textrm{N2}}$ is a transition to an incommensurate spiral state. ScMn$_6$Sn$_6$ lacks the collinear antiferromagnetic state.}
            \begin{tabular}{|c|c|c|c|c|c|}
                \hline
              \textit{R} & \textit{a} (\AA) & \textit{c} (\AA) & $T_{\textrm{N1}}$ (K) & $T_{\textrm{N2}}$ (K)
               & References\\ \hline
                    
             Sc& \begin{tabular}[c]{@{}c@{}}5.4649\\ 5.4692 \end{tabular} & \begin{tabular}[c]{@{}c@{}}8.9608\\ 8.9741\end{tabular} & \begin{tabular}[c]{@{}c@{}}390\\ 384 \end{tabular} & \begin{tabular}[c]{@{}c@{}}--\\ -- \end{tabular} & \begin{tabular}[c]{@{}c@{}}\cite{Richa2023}\\ \cite{magnetization}\\\end{tabular} \\ \hline
            
             Lu& \begin{tabular}[c]{@{}c@{}}5.5076 \\ 5.5079 \end{tabular} & \begin{tabular}[c]{@{}c@{}}8.9889\\ 8.9860\end{tabular} & \begin{tabular}[c]{@{}c@{}}395\\ 384 \end{tabular} & \begin{tabular}[c]{@{}c@{}}263\\ 220--250 \end{tabular} & \begin{tabular}[c]{@{}c@{}} This study \\ \cite{magnetization}\\\end{tabular} \\ \hline
             
             Y& \begin{tabular}[c]{@{}c@{}}5.541\\ 5.512\\ 5.5411 \end{tabular} & \begin{tabular}[c]{@{}c@{}}9.035\\ 8.984\\ 9.0228 \end{tabular} & \begin{tabular}[c]{@{}c@{}}345\\ 359\\ 340 \end{tabular} & \begin{tabular}[c]{@{}c@{}}--\\ 326\\ 333 \end{tabular} & \begin{tabular}[c]{@{}c@{}}\cite{Nirmal}\\ \cite{Li2021}\\\cite{YMn6Sn6_Rebecca}\end{tabular} \\ \hline
            \end{tabular}
        \end{table}

    \subsection{Magnetic Properties\label{chi}}
    Zero field neutron scattering experiments on LuMn$_6$Sn$_6$ and YMn$_6$Sn$_6$ show that below the N\'eel temperature $T_{\textrm{N1}}$, a commensurate collinear AF structure forms first with the propagation vector $k=(0, 0, 1/2)$~\cite{neutron1,neutron2,YMn6Sn_Heda_neutron,YMn6Sn6_Rebecca}. This is the magnetic configuration of phase AF1, which is the dominant phase in the (\textit{H},\textit{T}) phase diagram in Fig.~\ref{phsses}. 
    As the field increases in this phase, the moments gradually cant within the \textit{ab}-plane, forming the coplanar canted AF1 state~\cite{Hwan,YMn6Sn_Heda_neutron,YMn6Sn6_Rebecca}. The magnetic structures of the AF1 phase with and without an applied field are schematically shown in Fig.~\ref{crystal}(d).
    By cooling below $T_{\textrm{N2}}$, an incommensurate magnetic phase appears~\cite{neutron1,neutron2,LuMn6Sn6_LuMn6Ge6,magnetization,YMn6Sn_Heda_neutron,YMn6Sn6_Rebecca}. This phase has two nearly equal wave vectors and is known as a double-flat spiral or a double spiral. In this \textit{c}-axis helical order, the Mn moments in each bilayer rotate by a non-constant angle, requiring two distinct rotation angles to describe the directions of the moments~\cite{double_flat_spiral}. For YMn$_6$Sn$_6$, the resulting complex arrangement with a non-uniform rotation of the moments leads to a supercell structure with $c'=36c$~\cite{neutron2}. For LuMn$_6$Sn$_6$ this spiral phase corresponds to the AF2 phase as schematically shown in Fig.~\ref{crystal}(d).

    
    Upon application of an external magnetic field in the \textit{ab}--plane, several field-induced magnetic phases rise. In YMn$_6$Sn$_6$ the magnetic phases identified are a transverse conical spiral (TCS), fan-shaped (FL), and forced ferromagnet (FFM)~\cite{Nirmal,YMn6Sn6_Rebecca}. The cartoon in Fig.~\ref{crystal}(d) schematically shows these phases. 
    By comparing the ($H$, $T$) phase diagram in Fig.~\ref{phsses} with that of YMn$_6$Sn$_6$~\cite{YMn6Sn6_Rebecca,YMn6Sn6_Lifshitz}, it can be inferred that LuMn$_6$Sn$_6$ likely adopts a TCS and a FL magnetic structure in phases III and IV, respectively.

    Figure~\ref{M-R}(a) presents the temperature, $T$, dependence  of magnetization, $M$, obtained for a 0.05~T magnetic field applied along the \textit{a} and \textit{c} axis. The AF transition manifests itself as a kink in magnetization at $T_{\textrm{N1}} = 395$~K. A second transition, into an incommensurate spiral state, occurs at $T_{\textrm{N2}} = 263$~K, marking the boundary between the AF1 and AF2 magnetic phases. This latter transition appears more prominently as a drop in the $M(T)$ curve when the field is applied parallel to the \textit{a}-axis. The transitions temperatures are slightly higher than previously reported~\cite{Neel,magnetization}. 
    We observed that the use of higher purity tin and manganese results in a change of $T_{\textrm{N}}$'s to slightly higher values.
    There is a slight difference between the $T_{\textrm{N}}$'s values for the two orientations of the field. Our reported values are for measurements of $M~(T)$ in which the field is applied along the \textit{a}-axis. 
    
    The magnetic properties for the \textit{c}-axis magnetic field are shown in Fig.~S2 within SM~\cite{supp}. The magnetization curves for the field applied along the \textit{c}-axis show no field-induced transitions. This paper focuses on magnetotransport properties with an in-plane magnetic field. 
    
    When larger $a$-axis fields are applied, LuMn$_6$Sn$_6$ exhibits rich magnetism at low temperatures. Upon cooling below 200 K, it undergoes a series of magnetic phase transitions, as shown by the strong field dependence of $M$ in phases AF2$^*$, I,~II, and III in Fig.~\ref{M-R}(b). 
    Phase I was not clearly distinguishable in the field-dependence measurements of magnetization. Neutron scattering measurements by Do et al.~\cite{Hwan} show a continued change in the ordering wave vector below 250~K and a distinct kink at 100~K. Therefore, we labeled the magnetic phase below 100~K by AF2$^*$ to communicate this discontinuity in the character of the wave vector. 
    
    The AF1 state in YMn$_6$Sn$_6$~\cite{YMn6Sn6_2ndtrans_Hall,Li2021} exists in a very narrow temperature range, as can be seen from the small temperature difference between $T_{\textrm{N1}}$ and $T_{\textrm{N2}}$ in Table~\ref{table}. This phase is absent in ScMn$_6$Sn$_6$~\cite{Richa2023} which has a DS structure throughout the temperature range below its N\'eel temperature. However in LuMn$_6$Sn$_6$ the AF1 phase is much more extended in the phase diagram [Fig.~\ref{phsses}]. 
    Direct comparison of the magnetic phase diagrams of LuMn$_6$Sn$_6$, YMn$_6$Sn$_6$~\cite{YMn6Sn6_Rebecca,YMn6Sn6_Lifshitz}, and  ScMn$_6$Sn$_6$~\cite{ScMn6Sn6_APL}, reveals that the DS phase is divided into phases of I, II, AF2, and AF2$^*$ in the Lu compound. This is manifested in Fig.~\ref{M-R}(b) by the strong response of the helical AF2 state to the magnetic field at lower temperatures. The temperature derivative of the \textit{M}(\textit{T}) isotherm curves was used to construct the phase diagram, as shown by the square-shaped data points in Fig.~\ref{phsses}.

    \subsection{Electrical Transport Properties\label{resistivity}}
   \begin{figure*}[hbt]
		\includegraphics[width=1\linewidth]{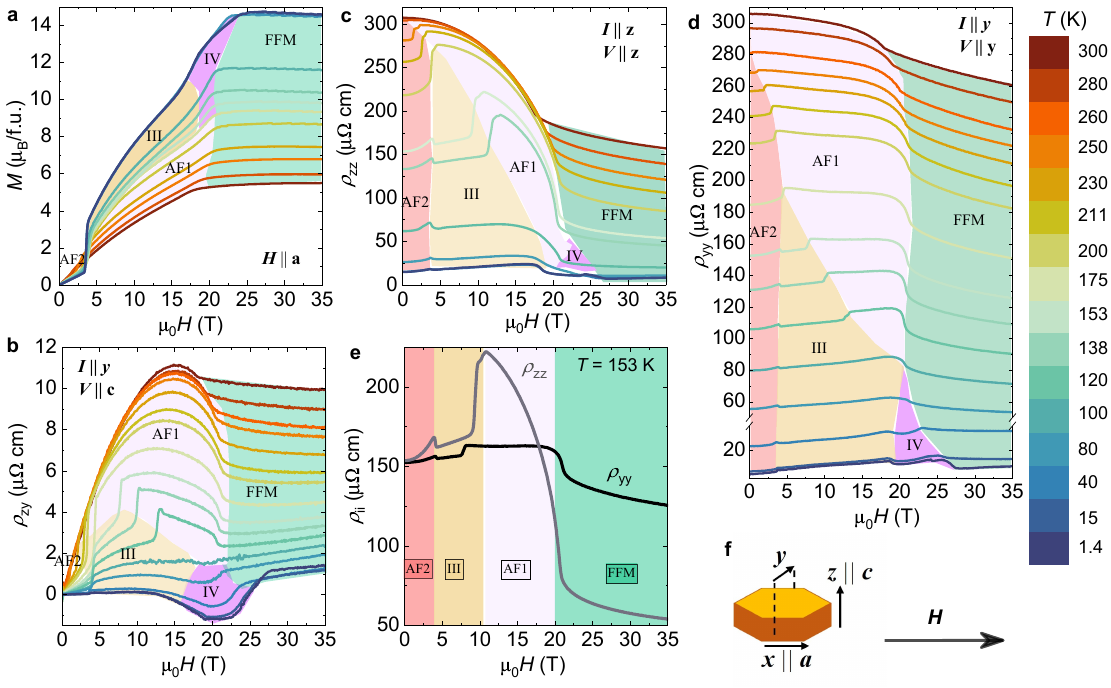}
		\caption{ Magneto-transport properties of LuMn$_6$Sn$_6$ for the magnetic field pplied along the \textit{a}-axis. (a) magnetization $M$, (b) Hall resistivity $\rho_{zy}$, (c) longitudinal resistivity $\rho_{zz}$, and (d) longitudinal resistivity $\rho_{yy}$ as a function of magnetic field at various temperatures. (e) Magnetoresistivities $\rho_{yy}$ and $\rho_{zz}$ show a strong anisotropy in the AF1 phase. The cartoon in (f) shows the directions of the applied electrical current, magnetic field, and the measured voltage.  The different magnetic phases are colored and labeled according to Fig.~\ref{phsses}. } \label{R-M-H}
	\end{figure*} 
     The intricate transport properties shown in Fig.~\ref{R-M-H} reveal several magnetic phases that are induced by the application of $\textbf{\textit{H}}~||~\textbf{a}$.  The field derivatives of these curves were used in constructing the ($H$,$T$) phase diagram in Fig.~\ref{phsses} [detailed in Fig.~S3 in SM~\cite{supp}].
     
    The \textit{M}(\textit{H}) isotherms in Fig.~\ref{R-M-H}(a) resemble those of \textit{R}Mn$_6$Sn$_6$ (\textit{R}= Sc and Y), albeit with the common magnetic phases appearing at much larger fields~\cite{Nirmal,ScMn6Sn6_APL,YMn6Sn6_Lifshitz,YMn6Sn6_emergent}.
    The abrupt change of $M$ around 4~T in Fig.~\ref{R-M-H}(a) (at AF2 to III phase boundary) suggests a first-order metamagnetic transition, which was seen in ScMn$_6$Sn$_6$ and YMn$_6$Sn$_6$ as well~\cite{Nirmal,YMn6Sn6_Pengcheng,ScMn6Sn6_APL}.  At low temperatures \textit{M} starts to saturate at $\mu_0H > 25$~T, while for ScMn$_6$Sn$_6$ and YMn$_6$Sn$_6$ the FFM phase appears at $\mu_0H >$ 3~T and 10~T, respectively~\cite{Nirmal,YMn6Sn6_Pengcheng,ScMn6Sn6_APL,YMn6Sn6_Lifshitz,YMn6Sn6_emergent}. The larger saturation field in LuMn$_6$Sn$_6$ indicates a stronger AF interaction between layers compared to the other sister compounds. 
    The $T$ dependence of the longitudinal resistivity in-plane ($\rho_{xx}$) and out-of-plane ($\rho_{zz}$) are shown in Fig.~\ref{M-R}(c). The electric current was applied along the crystallographic \textit{a}-axis (\textit{c}-axis) to measure $\rho_{xx}$ ($\rho_{zz}$). In Fig.~\ref{R-M-H}(b)--(e), the subscript $i$ in $\rho_{ji}$, indicates the direction of the applied current and $j$ denotes the direction of the measured voltage. A schematic of this measurement configuration is shown in Fig.~\ref{R-M-H}(f).  
    
    The transition temperatures appear as steps and more clearly in the temperature derivative of $\rho_{zz}$, Fig.~\ref{M-R}(c) inset.  $\rho_{zz}(T)$ shows a negative concavity in $T$ regions between $T_{N1}$ and $T_{N1}$. Often, the resistivity of an itinerant antiferromagnet shows positive concavity as a function of temperature in its ordered state. 
    This type of increase in electrical resistivity at the magnetic ordering temperature is commonly attributed to the formation of a superzone gap on the Fermi surface, where portions of the Fermi surface vanish, leading to an increase in resistivity~\cite{superzone_CeGe}. This is a result of the Fermi surface reconstruction caused by the emergence of a new magnetic Brillouin zone, which gaps out certain regions of the Fermi surface. The increase in $\rho_{zz}(T)$ upon lowering the temperature between $T_{N1}$ and $T_{N2}$ could also be a result of incoherent interlayer transport, perhaps due to spin scattering.
    
    Fig.~\ref{M-R}(c) shows that by further cooling below $T_{N2}$, $\rho_{zz}$ shows a positive concavity, observed in the whole $T$ range for $\rho_{xx}$, possibly due to the suppression of spin-disorder scattering. 
    We obtained residual resistivity ratios $RRR={R_{300K}}/{R_{2K}}=22-44$ and a residual resistivity of $RR= 4-15$ $\mu\Omega\textrm{cm}$ in different crystals.
	   
    Fig.~\ref{R-M-H}(c) and (d) show that both longitudinal resistivities  increase parabolically with $H$ in the AF2 phase and linearly in phase III.  In the AF1 phase $\rho_{zz}$ exhibit negative magnetoresistivity and $\rho_{yy}$  shows smaller changes.     
    Hall resistivity, shown in Fig.~\ref{R-M-H}(b), appears to be more sensitive to phase transitions around AF1 and III than magnetization, showing a sudden increase, as the magnetic structure evolves. These sharp increases in $\rho_{zy}$ are more pronounced at 120~K, 138~K, and 153~K and were consistently observed in multiple crystals and multiple batches. 
     
    At low fields, the field dependence of the Hall data have a positive slope, indicating that holes are the dominant charge carriers. We have obtained a rough estimate for the densities $n$ and mobilities $\mu$ of the holes by fitting $\rho_{zy} (H)$ to a simple band Drude model, which are shown in Fig.~S4 in SM~\cite{supp}. At room temperature, we obtained $n_{\textrm{300K}}=5.5\times10^{18}$~cm$^{-3}$ and $\mu_{\textrm{300K}}=3721$~cm$^2$V$^{-1}$S$^{-1}$, which drop to smaller values at lower temperatures. We also observed distinct Shubnikov-de Haas (SdH) oscillations in $\rho_{zz}$ at 1.4~K in the FFM phase for $\mu_0H>$27~T with the field aligned along the \textit{a}-axis. The SdH oscillations are shown in Fig.~S5 in SM~\cite{supp}. The analysis of the SdH signal, as explained in the SM~\cite{supp}, revealed a Fermi surface pocket with a $k$-space area of $A_k=7.2$~nm$^{-2}$. 

    \subsubsection{\textbf{Anomalous Hall Effect in the canted AF1 and FFM phases}}\label{Anomalous}
\begin{figure}[hbt]
		\includegraphics[width=0.9\linewidth]{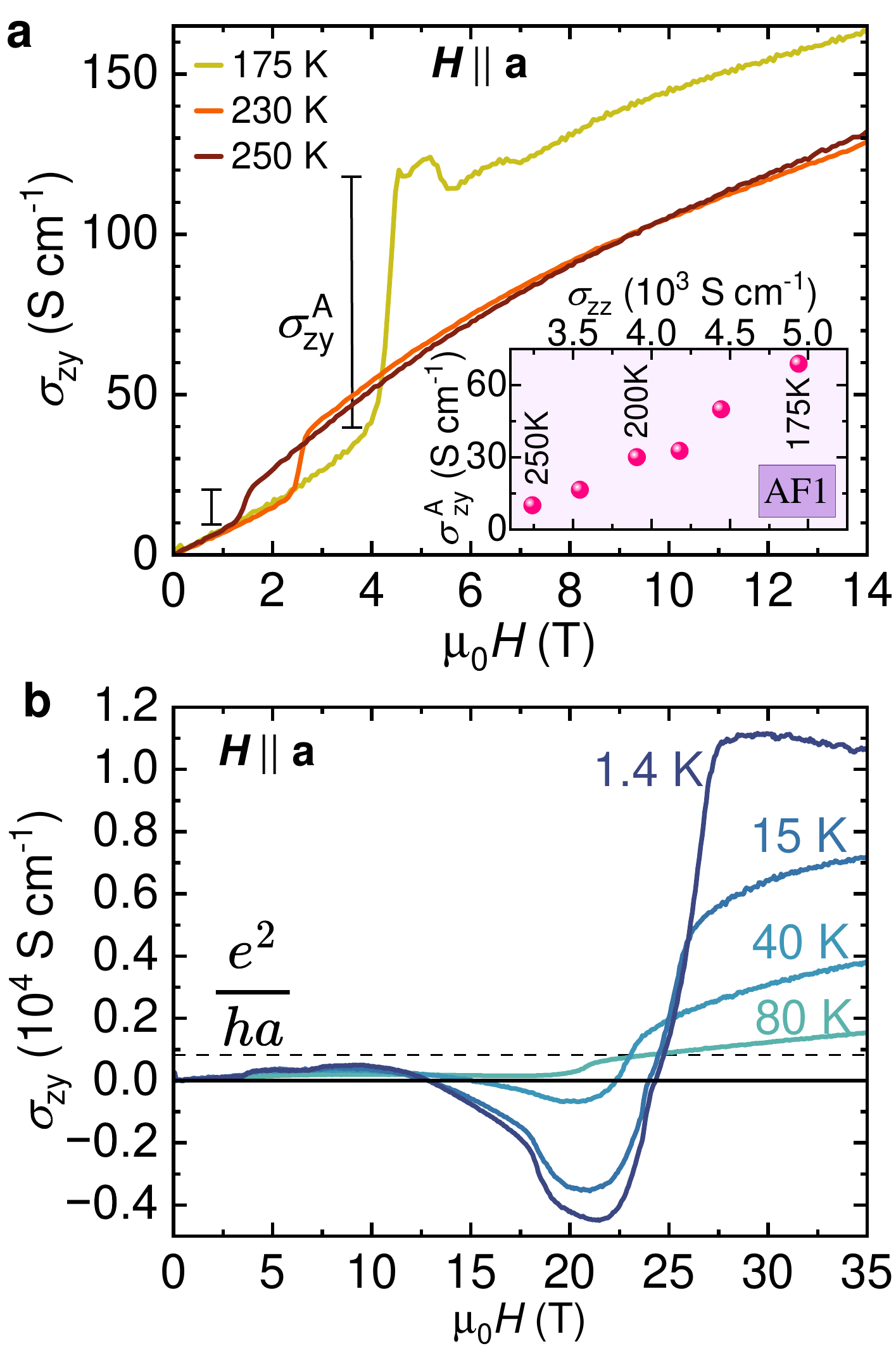}
		\caption{ Hall conductivity in LuMn$_6$Sn$_6$ as a function of magnetic field oriented along \textbf{\textit{a}}-axis at high temperatures in the canted AF1 phase (a) and low temperatures (b). The vertical lines in (a) show the extracted anomalous contribution $\sigma_{zy}^A$ to the Hall conductivity in the canted AF1 phase, which are plotted against conductivity in the inset. The upper threshold value of the intrinsic AHC in three-dimensions ($e^2/h\textit{a}$) is indicated by the dashed line in (b). 
        } \label{conductivity}
	\end{figure}

    We further explore the Hall effect data by examining the conductivity $\sigma_{zz}$ and Hall conductivity $\sigma_{zy}$, shown in Fig.~\ref{conductivity}. These curves were obtained by inverting the resistivity tensor~\cite{kittel}, following Eq.~(1). 
    \begin{equation}
		\begin{bmatrix}
                \sigma_{yy} & \sigma_{yz} \\
                \sigma_{zy} & \sigma_{zz}
        \end{bmatrix} = 
        \begin{bmatrix}
                \rho_{yy} & \rho_{yz} \\
                \rho_{zy} & \rho_{zz}
        \end{bmatrix}^{-1}=\frac{1}{\rho_{yy}\rho_{zz}+\rho_{zy}^2}
        \begin{bmatrix}
                \rho_{zz} & -\rho_{yz} \\
                -\rho_{zy} & \rho_{yy}
        \end{bmatrix}
	\end{equation}\label{Eq:resistivity_tensor}
     
     As Fig.~\ref{conductivity}(a) shows, there is a sudden increase in the Hall conductivity at field values where the metamagnetic transitions in the magnetization [Fig.~\ref{R-M-H}(a)] occur.  This happens at the boundary between the AF2 and AF1 phases. At these field strengths, the canting of magnetic moments along the applied field introduces an effective ferromagnetic component, which gives rise to an anomalous contribution to the Hall conductivity [see the schematics for the canting of the moments in Fig.~\ref{crystal}(d)]. Note that canting of the magnetic moments happens in the plane~\cite{Hwan,YMn6Sn_Heda_neutron,YMn6Sn6_Rebecca}. The anomalous Hall conductivity (AHC) in the AF1 phase, $\sigma_{zy}^A$, is estimated directly from the sharp increase in $\sigma_{zy}$ values marked by the black vertical lines in Fig.~\ref{conductivity}(a). These values are plotted in the inset of Fig.~\ref{conductivity}(a) along with the conductivity values. 
      
     The intrinsic AHE, which is related to the Berry curvatures in momentum space theoretically should be less than $e^2/ha$~\cite{AHE_Allan,AHE_MnGe}. Here, $e$ is the elementary charge, $h$ is Planck's constant, and $a$ is the typical lattice parameter, which for LuMn$_6$Sn$_6$ would be $\approx 800$ S cm$^{-1}$.      
     The extracted $\sigma_{zy}^A$ values in the inset of Fig.~\ref{conductivity}(a) are within the limit of the intrinsic AHC. Therefore, we conclude that the sudden increase in the $\rho_{zy}$ values upon entering phase AF1 is due to intrinsic contributions to the AHE. The maximum value of $\sigma_{zy}^A$ that we can unambiguously extract is about 70~S~cm$^{-1}$ at 175 K. This is less than the typical values for intrinsic AHC in ferromagnets such as Fe ($>1000$ S cm$^{-1}$)~\cite{AHC_Fe} or in  itinerant \textit{d}-electron AF materials such as Mn$_3$Sn, Mn$_3$Ge, and NbMnP (100--450 S cm$^{-1}$)~\cite{AHE_Mn3Sn,AHE_Mn3Ge,AHE_NbMnP}. 

     Another possible explanation for the sudden increase in the Hall effect data is the metamagnetic multiband Hall effect~\cite{ErGa2_MultiBandHall}. In this scenario, the non-monotonic character of the Hall effect is attributed to a band-dependent mobility modulation resulting from 4\textit{f}-5\textit{d} interactions. The Kondo-like Hamiltonian in this model is particularly effective at low temperatures and low fields, while the sharp jump in $\rho_{zy}$ in LuMn$_6$Sn$_6$ occurs at high fields and relatively elevated temperatures. Therefore, we conclude that the sharp increase in the Hall conductivity at the border of the AF2 to AF1 phase is related to the formation of a ferromagnetic component as the result of canting of the moments. 

     We now turn our attention to the low-temperature transport properties. Scaling relations between longitudinal and Hall conductivities are often employed empirically to identify the dominant scattering mechanisms that contribute to anomalous transport~\cite{AHE_Allan}. The moderate longitudinal conductivity of $\sigma_{zz} (T=1.4$~K) in the order of $10^5$~S~cm$^{-1}$ 
     places LuMn$_6$Sn$_6$ in the moderate conductivity regime where intrinsic scattering phenomena dominate the AHE. 
     However, the $\sigma_{zy}$ values increase sharply at low $T$'s when the field is ramped into the FFM phase, as shown in Fig.~\ref{conductivity}(b). In fact, Hall conductivity at 1.4~K reaches 11,000~S~cm$^{-1}$, exceeding the intrinsic quantized AHE of $e^2/ha$ by two orders of magnitude. This shows a clear deviation from the conventional scaling relations established for ferromagnets. Such large AHE with electron-scattering origin (extrinsic mechanisms--such as skew scattering) has been reported in the chiral ferromagnetic state of MnGe thin films~\cite{AHE_MnGe}. 
     
     Equally interesting is the sudden change in the sign of $\sigma_{zy}$ near 25~T, at the boundary of phase IV(V) to FFM phase. Below 25~T, the Hall conductivity is dominated by electron bands. However, once the field exceeds 25~T, $\sigma_{zy}$ becomes positive, indicating a shift in dominant charge carriers from electrons to holes.
     By directly comparing the magnetic phase diagram of LuMn$_6$Sn$_6$ (Fig.~\ref{phsses}) with YMn$_6$Sn$_6$~\cite{YMn6Sn6_Rebecca,YMn6Sn6_Lifshitz}, it can be inferred that the magnetic structure undergoes a transition from a fan-like (FL) phase to the FFM phase around 25~T. The observed sign change in $\sigma_{zy}$ may be related to a magnetic field-induced reconstruction of the Fermi surface. In particular, a magnetization-driven Lifshitz transition has been reported for YMn$_6$Sn$_6$~\cite{YMn6Sn6_Lifshitz}, suggesting a similar mechanism could be operative in LuMn$_6$Sn$_6$.    
     Our results suggest the need for further theoretical studies on the electronic band structure of LuMn$_6$Sn$_6$, the enhancement of Berry curvature, and its relationship to the observed AHE at these field-induced magnetic phases. 


    \subsubsection{\textbf{Pitfalls in the extraction of THE and the role of resistivity anisotropy}}\label{Pitfalls}
    In this section, we demonstrate the limitations of standard methods used to extract the topological Hall effect (THE). To do so, we first apply these empirical techniques to the AF1 phase, which is not expected to exhibit any THE. This allows us to assess which method yields results most consistent with theoretical expectations. In the following section, we then apply the method that shows the best agreement to phase III (the TCS phase), which is reported to show a THE in related compounds such as YMn$_6$Sn$_6$~\cite{Nirmal,YMn6Sn6_Pengcheng} and ErMn$_6$Sn$_6$~\cite{Er_fazel}.
       
    \begin{figure}[hbt]
		\includegraphics[width=1\linewidth]{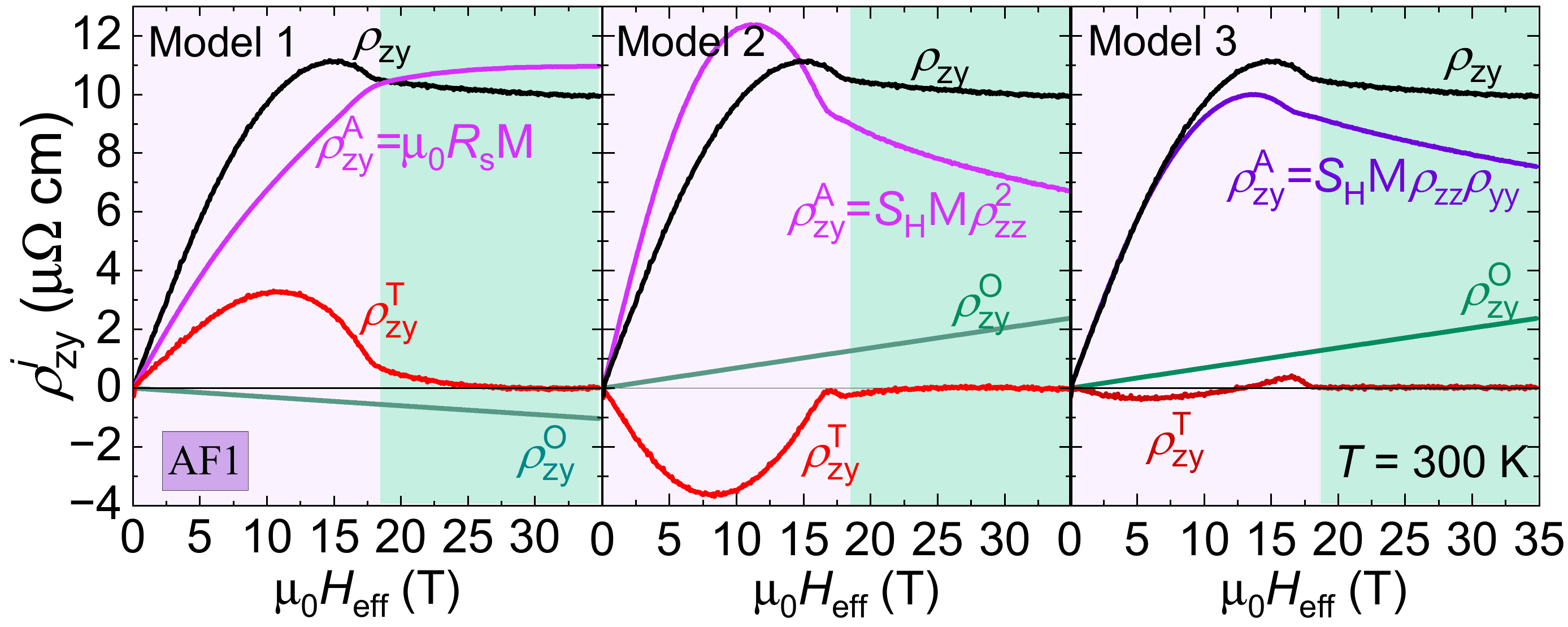}
		\caption{
        Comparison of empirical approaches for modeling the AHE in the extraction of the THE. Panels display the field dependence of Hall resistivity $\rho_{zy}$ and its components: ordinary ($\rho^{O}_{zy}$), anomalous ($\rho^{A}_{zy}$), and topological ($\rho^{T}_{zy}$) for LuMn$_6$Sn$_6$ at 300~K. $\rho^{A}_{zy}$ is estimated using three empirical models. In model~1, $\rho^{A}_{zy}$ is assumed to be linearly proportional to the magnetization. In models~2 and 3, $\rho^{A}_{zy}$ is taken to be proportional to the magnetization multiplied by the square of the longitudinal resistivity.
        Model~2 assumes isotropic longitudinal resistivity setting  $\rho_{yy} = \rho_{zz}$, leading to an exaggerated value for $\rho^{T}_{zy}$. In model~3, $\rho^{A}_{zy}$ is derived by accounting for the anisotropy in the longitudinal resistivity. Note that $\chi_{ijk}=0$ for the AF1 phase, so no THE is expected.} \label{THE_False}
	\end{figure}

     The Hall effect behavior in Fig.~\ref{R-M-H}(b) is quite intriguing. At high temperatures, $\rho_{zy}$ shows a hump within the AF1 phase that could mistakenly be regarded as resulting from a THE contribution. The three main contributions to the Hall effect are the ordinary Hall effect $\rho_{zy}^{O}$, the anomalous Hall resistivity $\rho_{zy}^{A}$, and the topological Hall resistivity $\rho_{zy}^{T}$; $\rho_{zy}=\rho_{zy}^{O}+\rho_{zy}^{A}+\rho_{zy}^{T}$. 
    The THE is extracted by first fitting the high field Hall data--where only the ordinary and anomalous Hall effects are present--to isolate their contributions. This fitted background is then subtracted from the total Hall signal, and the remaining part is attributed to the topological Hall effect. The process of extracting the topological Hall resistivity is detailed in section S~6 of SM~\cite{supp} for a representative temperature of 40~K.
    
    The ordinary Hall effect is approximately linear with respect to the applied field, $\rho_{zy}^{O}=R_0B$~\cite{kittel}. Where, $R_0$ is the ordinary Hall coefficient and $B$ is the induction field. In a multi-band system, $R_0$ can be field dependent. 
    
    Two empirical equations are frequently employed in the experimental analysis of the AHE~\cite{Nirmal,YMn6Sn6_Pengcheng,FeGeTe_THE,ScMn6Sn6_APL,Er_fazel}. One of them, derived for ferromagnets, is given by $\rho_{zy}^{A}=\mu_0R_sM$, where $\mu_0$ is the vacuum permeability and $R_s$ is the anomalous Hall coefficient. This relation for $\rho_{zy}^{A}$ is used in model~1 in Fig.~\ref{THE_False} for the extraction of $\rho_{zy}^{T}$. Using this empirical relation for the AHE results in an apparent topological Hall resistivity that peaks at 3~$\mu\Omega$~cm at 300~K. 
    
    Another commonly used relation is expressed as $\rho_{zy}^{A}= R_sM\rho_{zz}\rho_{yy}$, where $\rho_{zz}(\rho_{yy})$ and $\rho_{zy}$ correspond to longitudinal and Hall resistivities, respectively. Here, $\rho_{zy}^{A}$ is derived using Eq.~(1) and assuming that anomalous Hall conductivity is proportional to magnetization. We refer to this equation for the AHE as model~3. 
    
     In most metallic systems, longitudinal resistivities are isotropic, making it reasonable to express $\rho_{zy}^{A}=R_sM\rho_{yy}^2$. However, in non-cubic compounds with anisotropic resistivity, especially in out-of-plane measurements (\textit{zy}-plane), if both longitudinal resistivities are not carefully considered, as shown in Fig.~\ref{THE_False} by model~2, the resulting AHE is overestimated. This overestimation of $\rho_{zy}^{A}$ consequently leads to an exaggerated value of 4~$\mu\Omega$~cm for $\rho_{zy}^{T}$, which is larger than the values reported for well-established skyrmionic systems such as MnSi~\cite{THE_MnSi} and Gd$_2$PdSi$_3$~\cite{Gd2PdSi3_THE}.
     
    Why does accounting for the anisotropy of magnetoresistance make such a big difference in LuMn$_6$Sn$_6$? As shown in Fig.~\ref{R-M-H}(c) and (d), in the canted AF1 phase, $\rho_{zz}$ exhibits a strong negative magnetoresistivity, while $\rho_{yy}$ curves remain relatively flat.
    The strong anisotropy in the longitudinal resistivities in the AF1 phase is depicted in Fig.~\ref{R-M-H}(e) for a representative temperature of 153~K. This pronounced anisotropy in the magnetoresistivities plays a critical role in the proper analysis of the Hall effect data. This can be seen by comparing panels in model~2 and model~3 in Fig.~\ref{THE_False}.      
    In model~3, we have now included the anisotropy in the magnetoresistivity, however, we still obtain a non-zero topological Hall effect at 300~K, where the magnetic structure is coplanar canted antiferromagnet with zero scalar spin chirality--conditions under which a THE is not expected. 

    As demonstrated in this section (and also in the next one), there are limitations in using these experimental tools to analyze the contribution from the AHE.  
    Attention must also be paid to the procedure for separating the contribution from the ordinary Hall effect. In systems where there is a possibility of a change in the Fermi surface at different magnetic structures such as Lifschitz transitions~\cite{YMn6Sn6_Lifshitz}, $R_0$ is no longer a constant, and the contribution from the ordinary Hall effect cannot be assumed to be equal in all magnetic phases. At minimum, we recommend using model~3 to estimate THE to approximately estimate the anomalous Hall contribution in anisotropic phases. 
    The need for a more thorough consideration in identifying emergent magnetic fields in frustrated itinerant magnets has recently been discussed by Refs.~\cite{ErGa2_MultiBandHall,THE_Challanges} as well.
    

     \subsubsection{\textbf{Topological Hall Effect in phase III (transverse conical spiral, TCS)}} 
     
    We now turn our attention to the magnetic phase III and the possibility of observing the THE in this phase. THE is a hallmark of scalar spin chirality, and is observed in skyrmionic lattices such as MnSi~\cite{THE_MnSi}, FeGe~\cite{THE_FeGe}, Gd$_2$PdSi$_3$~\cite{Gd2PdSi3_THE}, and EuAl$_4$~\cite{THE_EuAl4}.          
    By directly comparing the magnetic phase diagrams of LuMn$_6$Sn$_6$ and YMn$_6$Sn$_6$~\cite{YMn6Sn6_Rebecca,YMn6Sn6_Lifshitz} along with neutron scattering studies reported in Ref.~\cite{Hwan}, it can be concluded that the magnetic structure of this phase is a transverse conical spiral (TCS). A THE is not expected in this phase due to its zero scalar spin chirality~\cite{YMn6Sn6_Rebecca,Nirmal}. However, the TCS phase in YMn$_6$Sn$_6$ has been reported to exhibit a THE~\cite{Nirmal,YMn6Sn6_Pengcheng}. The THE occurs at high temperatures, leading to the proposal of a dynamical mechanism as a possible explanation for the appearance of THE in the TCS phase, despite the absence of scalar spin chirality.
    It has been suggested that magnon fluctuations, coupled with the strongly two-dimensional nature of the magnetic exchange, may explain the observed THE in YMn$_6$Sn$_6$~\cite{Nirmal}. 
    Following this observation, we examine the THE in the TCS phase of LuMn$_6$Sn$_6$. 
     
\begin{figure}[hbt]
	\includegraphics[width=0.8\linewidth]{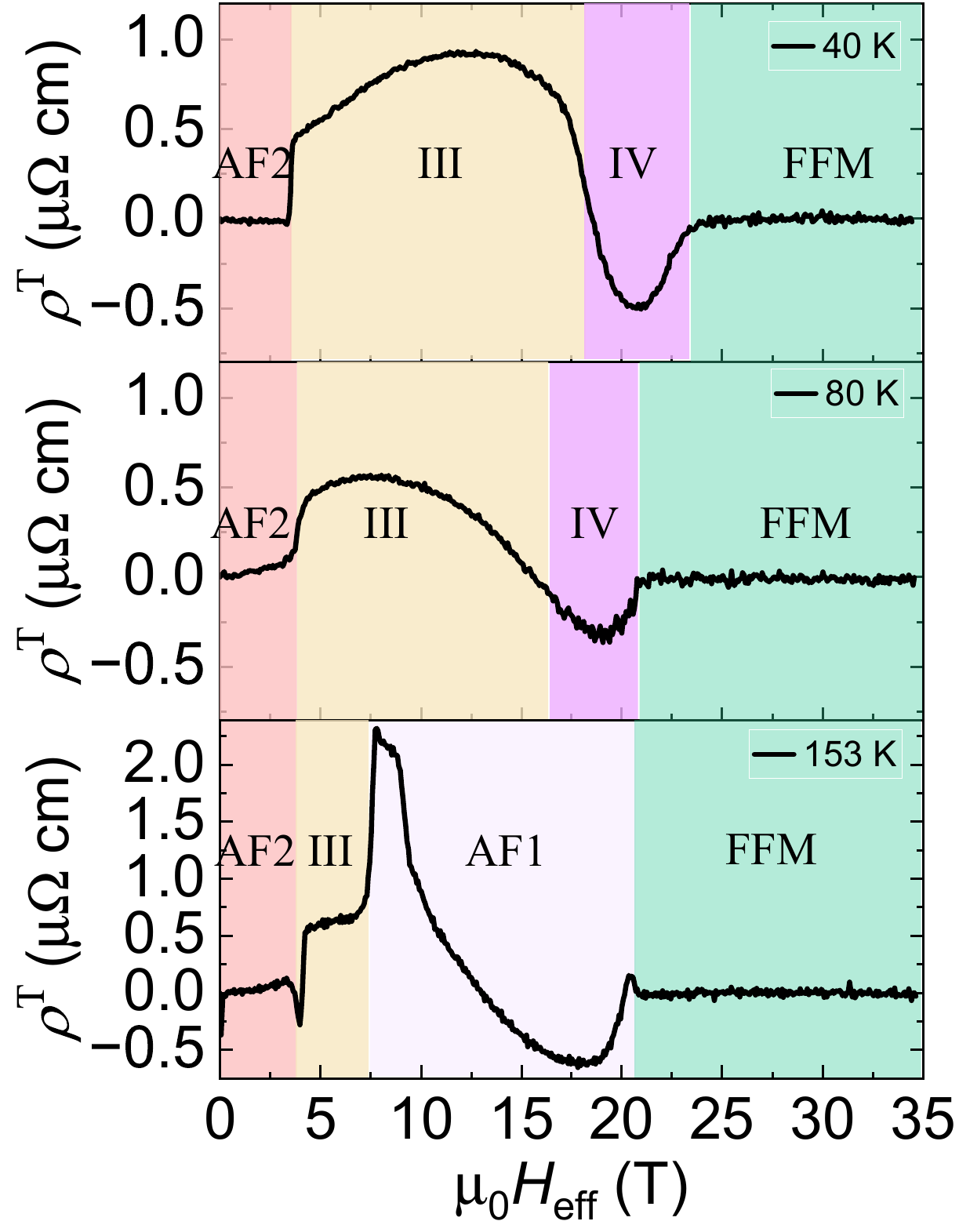}
	\caption{Extracted topological Hall $\rho^T$ at several constant temperatures. The anomaly in $\rho^T$ across the border of phase III and AF1 at 153~K, is a result of an artifact due to mismatch between the transition fields in $\rho_{yy}$ and $\rho_{zz}$. This mismatch can be see in Fig.~\ref{R-M-H}(e) around 10~T. 
    The observation of THE at phase AF1 with $\chi_{ijk}=0$ calls into question our standard methods for extracting the THE. } \label{THE}
    \end{figure}

   The derived topological contribution to the Hall resistivity at 40, 80, and 153~K are plotted as a function of effective magnetic field in Fig.~\ref{THE} (See Fig.~S6 in SM~\cite{supp} for detailed calculations for a representative temperature of 40~K). We were unable to clearly separate the contributions of $\rho_{zy}^{A}$ and $\rho_{zy}^{T}$ to $\rho_{zy}$ at temperatures below 40~K.
   In our analysis, for the AHE we used the empirical relation from model~3, $\rho_{zy}^{A}= R_sM\rho_{zz}\rho_{yy}$, as it provided slightly better agreement with the expected absence of a THE in the AF1 phase (Sec.~\ref{Pitfalls}).
   
     After properly accounting for the anisotropy of the longitudinal resistivity, we obtained a maximum value of 0.93~$\mu\Omega$~cm at 40~K for $\rho_{zy}^{T}$ in phase III (the TCS phase). This is within the reported value of 1~$\mu\Omega$~cm for the TCS phase of YMn$_6$Sn$_6$ and ScMn$_6$Sn$_6$ at 245~K~\cite{Nirmal,YMn6Sn6_Pengcheng,ScMn6Sn6_APL}. We note that the value of THE would be significantly larger if we had used model~2 for the estimation of the AHE (as we have emphasized in Sec.~\ref{Pitfalls}).

    $\rho_{zy}^{T}$ in phase III exhibits slightly lower values at elevated temperatures of 80 and 153~K. A non-vanishing THE is also observed in phase IV, which is most likely associated with a FL magnetic structure~\cite{Hwan}.  Particularly striking is the presence of a non-zero THE at 153~K in the AF1 phase. Given that AF1 is characterized by a canted antiferromagnetic structure with no scalar spin chirality, this observation is unexpected.     
    It is concerning that THE is present in all three phases--III, AF1, and IV--with nearly comparable magnitudes. Notably, the THE in phase III is exceptionally robust, persisting across a wide magnetic field range of up to 20~T.
    
    The observations in this section and Sec.~\ref{Pitfalls} raise questions about the reliability of our conventional methods for extracting THE. Our observations cast doubt on whether the TCS phase truly exhibits a topological Hall signal. As discussed in Sec.~\ref{Pitfalls}, there are limitations in using the empirical relations for partitioning the Hall effect data when interpreting topological transport data. The peculiar behavior of THE in $R$Mn$_6$Sn$_6$ suggests the need for further theoretical studies to understand the origin of the emerging THE and AHE, and more thorough consideration when dealing with anomalies in the Hall effect data.

	\section{Conclusion}\label{Conclusion}
    Our detailed magnetotransport study of the metallic kagome compound LuMn$_6$Sn$_6$ reveals a rich magnetic phase diagram driven by competing interactions 
    under an applied in-plane magnetic field. 
    Our investigation of the Hall effect in the canted antiferromagnetic (AF1) state, shows an anomalous Hall effect (AHE) which we attribute to intrinsic contributions. We measured a Hall conductivity of 11,000~Scm$^{-1}$ at 1.4~K, which exceeds the intrinsic quantized AHE of $e^2/ha$ by two orders of magnitude, despite moderate longitudinal conductivity $\sigma_{zz}$ on the order of $10^5$~S~cm$^{-1}$.
    
    We also demonstrate the limitations of standard methods used to estimate the various contributions to the Hall effect. These limitations can give rise to a large apparent topological Hall effect (THE) in coplanar magnetic phases where a true topological Hall response is not anticipated. This issue is exacerbated when the anisotropy in the longitudinal resistivity is not adequately accounted for. Furthermore, even when resistivity anisotropy is considered, a residual THE-like signal persists in phases III, IV, and AF1, despite their lack of scalar spin chirality. 
    These findings highlight the need for more rigorous theoretical frameworks to accurately interpret the Hall effect data in complex magnetic compounds.


    

	\section*{Acknowledgment}
	S.~M., R.~P.~M., W.~R.~M., and D.~M. acknowledge the support from the Gordon and Betty Moore Foundation's EPiQS initiative, Grant GBMF9069 and the support from AFOSR MURI Grant No. FA9550-20-1-0322. This publication is funded in part by a QuantEmX grant from ICAM and the Gordon and Betty Moore Foundation through Grant GBMF9616 to Shirin Mozaffari. L.~B. is supported by the US Department of Energy (DOE) through the BES program, award DE-SC0002613.
    X-ray diffraction was performed at the Institute for Advanced Materials $\&$ Manufacturing (IAMM) Diffraction facility, located at the University of Tennessee, Knoxville.
    The National High Magnetic Field Laboratory is supported by the National Science Foundation through NSF/DMR-1644779 and NSF/DMR-2128556 and the State of Florida.
	
	\hfill
	
	\section*{References}
	
	\bibliography{Lu166}

\end{document}


\title{Robust Topological Hall in commensurate phase of LuMn$_6$Sn$_6$}
	
		\author{Shirin Mozaffari}
	\email{smozaff1@utk.edu}
	\affiliation{Department of Materials Sciences and Engineering, The University of Tennessee, Knoxville, TN 37996, USA}

   \author{Seung-Hwan Do}
      \affiliation{Department of Materials Sciences and Engineering, The University of Tennessee, Knoxville, TN 37996, USA}
 
	\author{Richa Madhogaria}
	\affiliation{Department of Materials Sciences and Engineering, The University of Tennessee, Knoxville, TN 37996, USA}

	\author{Aikaterini Flessa Savvidou}
	\affiliation{National High Magnetic Field Laboratory, Florida State University, Tallahassee, Florida 32310, USA}

	\author{Brian Casas}
	\affiliation{National High Magnetic Field Laboratory, Florida State University, Tallahassee, Florida 32310, USA}
 
        \author{William R. Meier}
	\affiliation{Department of Materials Sciences and Engineering, The University of Tennessee, Knoxville, TN 37996, USA}
  
	\author{Rui Xue}
	\affiliation{Department of Physics and Astronomy, The University of Tennessee, Knoxville, TN 37996, USA}
	
	\author{Eun Sang Choi}
	\affiliation{National High Magnetic Field Laboratory, Florida State University, Tallahassee, Florida 32310, USA}
	
	\author{Luis Balicas}
	\affiliation{National High Magnetic Field Laboratory, Florida State University, Tallahassee, Florida 32310, USA}
	
	\author{David G. Mandrus}
	\email{dmandrus@utk.edu}
	\affiliation{Department of Materials Sciences and Engineering, The University of Tennessee, Knoxville, TN 37996, USA}
	\affiliation{Department of Physics and Astronomy, The University of Tennessee, Knoxville, TN 37996, USA}

\maketitle

\renewcommand{\thesection}{S\arabic{section}}
	\section{Powder X-ray Diffraction \label{Powder X-ray Diffraction}}
\begin{figure}[htb]
		\renewcommand{\thefigure}{S\arabic{figure}}
		\begin{center}
			\includegraphics[width = 0.5\linewidth ]{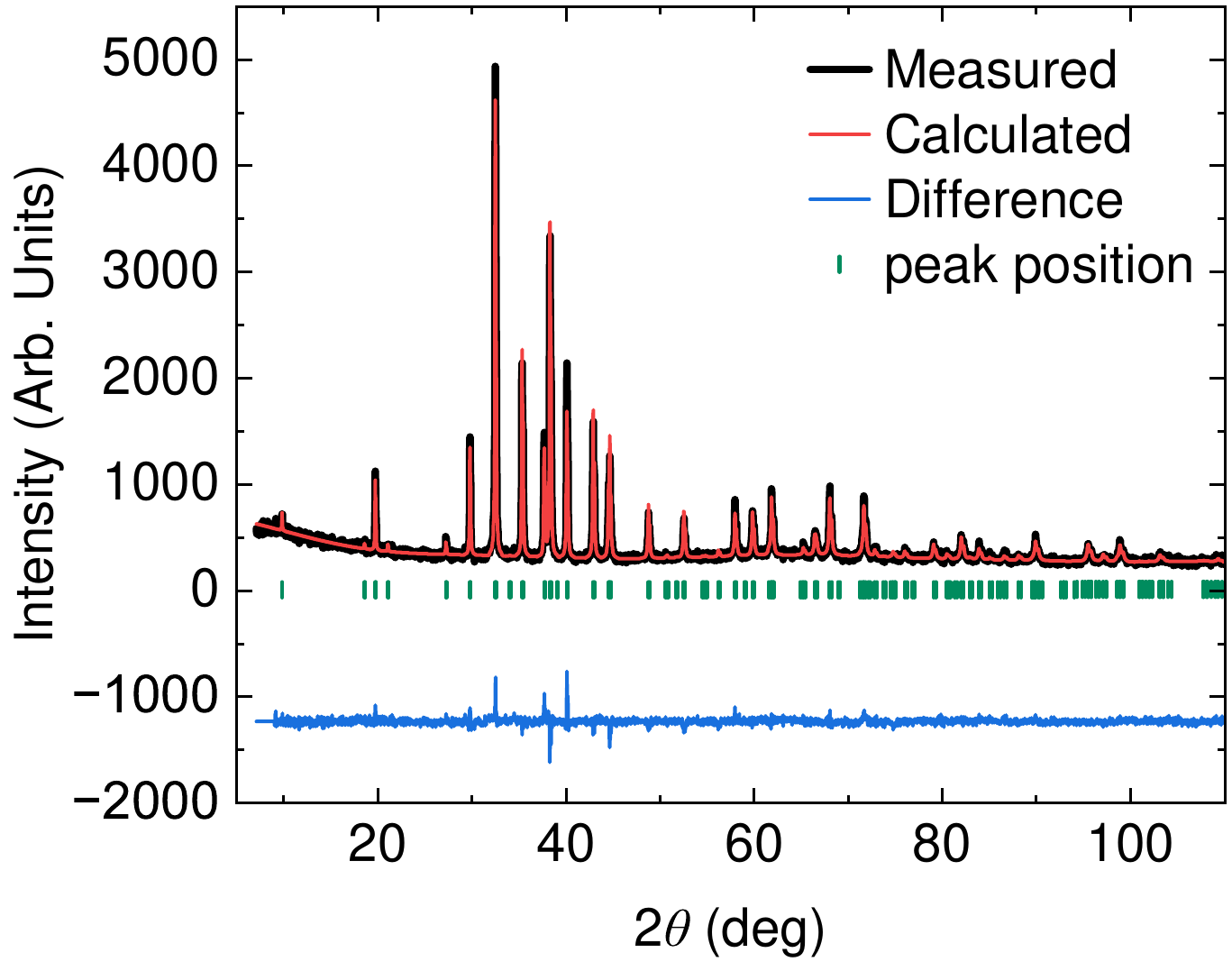}
			\caption{Powder x-ray diffraction refinement of crushed LuMn$_6$Sn$_6$ single crystals using Cu $K_{\alpha}$ radiation.} \label{XRD}
		\end{center}
	\end{figure}

	Powder X-Ray Diffraction measurements were performed on ground crystals in a Panalytical Empyrean diffractometer with Cu \textit{K}$_\alpha$ X-ray source. Refinement with FullProf~\cite{Fullprof}, Figs.~\ref{XRD}~(a) revealed that LuMn$_6$Sn$_6$ has a \textit{P6/mmm} HfFe$_6$Ge$_6$-type structure. 

\newpage

\renewcommand{\thesection}{S\arabic{section}}
	\section{Magnetic susceptibility and Hall effect for field applied out-of-plane\label{Magnetic susceptibility}}
\begin{figure}[htb]
		\renewcommand{\thefigure}{S\arabic{figure}}
		\begin{center}
			\includegraphics[width = 0.8\linewidth ]{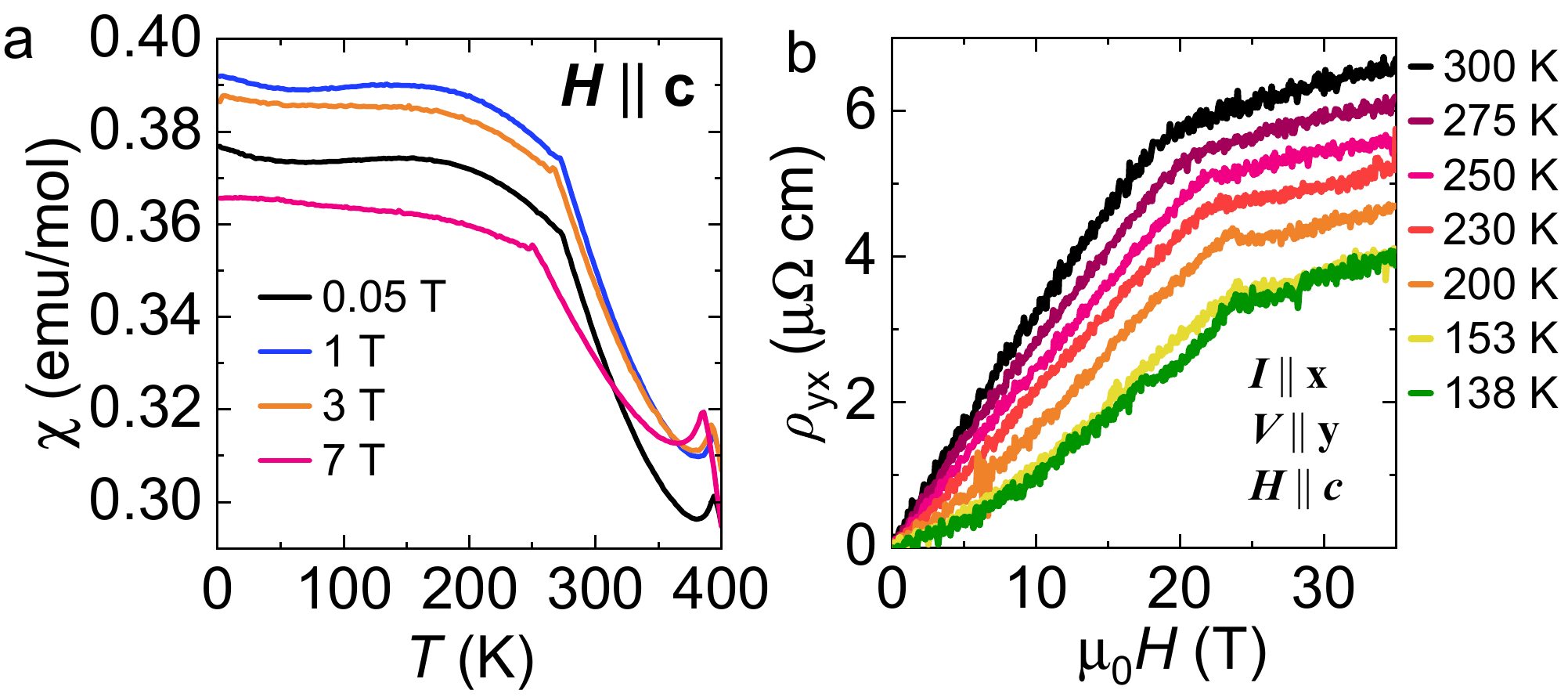}
			\caption{(a) Temperature dependence of the magnetic susceptibility, for applied field $\mu_0H$ along the \textit{c}-axis for LuMn$_6$Sn$_6$, obtained on heating the sample after zero-field-cooling. (b) Hall resistivity $\rho_{yx}$ as a function of magnetic field that was applied along the crystalline \textit{c} direction at several temperatures.} \label{M-T_field_par_c}
		\end{center}
	\end{figure}

	Magnetic susceptibility ($\chi$) in Fig.~\ref{M-T_field_par_c}(a) reveals two antiferromagnetic phase transitions occurring at $T_{\textrm{N1}}=394$~K and $T_{\textrm{N2}}=273$~K at 0.05~T. These transition temperatures change by the application of stronger field and are slightly different from those in which the field was applied parallel to the \textit{a}-axis [Fig~3(a)]. 
    
    Figure ~\ref{M-T_field_par_c}(b) shows the field dependence of Hall resistivity. In this configuration, the electric current was applied in the plane and the magnetic field was applied along crystalline \textit{c}-axis. 
    As can be seen, the magnetic moments gradually cant toward the field. The Hall data for the out-of-plane field show no interesting magnetic phase transitions that were observed for the in-plane field.

\newpage

\renewcommand{\thesection}{S\arabic{section}}
	\section{Temperature  and magnetic field phase diagram\label{phases_explain}}
\begin{figure*}[htb]
		\renewcommand{\thefigure}{S\arabic{figure}}
		\begin{center}
			\includegraphics[width = \linewidth ]{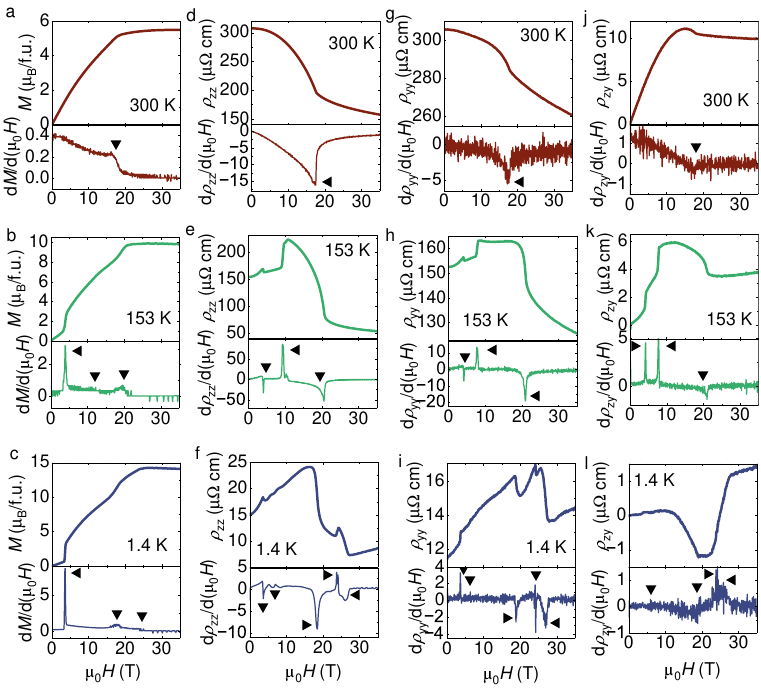}
			\caption{(a), (b), are (c) Magnetization at 300~K, 153~K, and 1.4~K, respectively, as a function of magnetic field. (d), (e), and (f) are the field dependence of out-of-plane longitudinal resistivity, $\rho_{zz}$,at 300~K, 153~K, and 1.4~K, respectively. (g), (h), and (i) are the field dependence of in-plane longitudinal resistivity, $\rho_{yy}$,at 300~K, 153~K, and 1.4~K, respectively. (j), (k), and (l) are the field dependence of Hall effect resistivity, $\rho_{zy}$,at 300~K, 153~K, and 1.4~K, respectively.
            The corresponding field derivatives is displayed below each plot. Triangles mark the position of magnetic phase transition on the derivatives curves.
            }
            
            \label{phases_explained}
		\end{center}
	\end{figure*}

    The derivative of magnetization and electrical transport data with respect to magnetic field was used to construct the magnetic phase diagram in Fig.~1 in the main text. This is shown for a few representative temperatures in Fig.~\Ref{phases_explained}. Similar procedure was performed for the temperature derivative of $M(T)$ curves. 

\newpage

\renewcommand{\thesection}{S\arabic{section}}
	\section{Charge carrier densities and mobilities \label{carriers}}
	A rough estimation for the charge carrier densities $n$ and mobilities $\mu$ for holes was performed by fitting low-field Hall resistivity and magnetoresistivity data to the Drude model. The results are shown in Fig.~\ref{carriers}. The positive slope of the Hall resistivities data at low fields [Fig.~4(b)] indicates that hole-type carriers dominate the transport. The slope of the Hall resistivity becomes smaller at lower temperatures. 

    	\begin{figure}[htb]
		\renewcommand{\thefigure}{S\arabic{figure}}
		\begin{center}
			\includegraphics[width = 0.45\linewidth ]{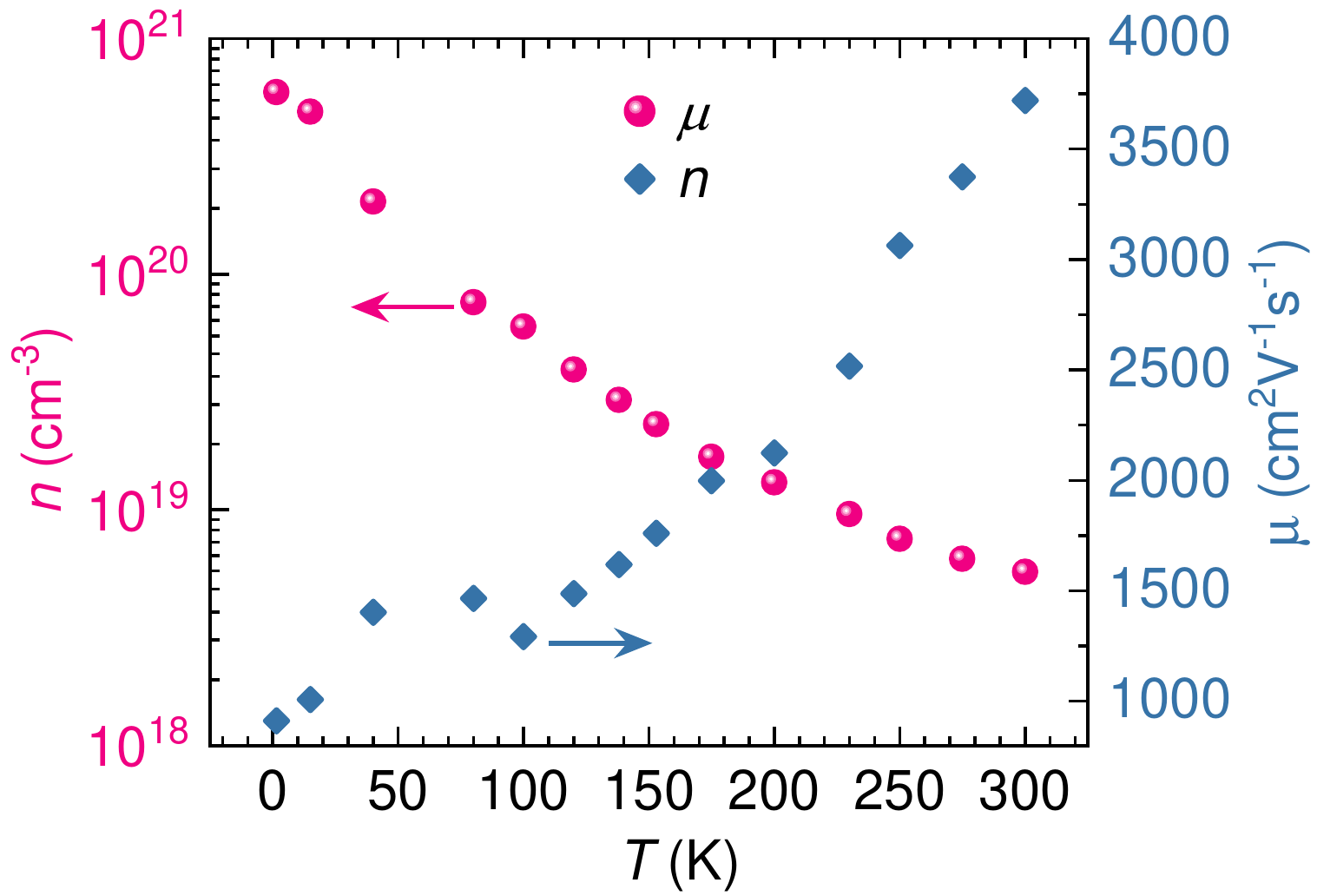}
			\caption{A rough estimation of the charge carrier densities $n$ and mobilities $\mu$ for holes as a function of temperature.} \label{carriers}
		\end{center}
	\end{figure}

\renewcommand{\thesection}{S\arabic{section}}
	\section{Quantum Oscillation \label{SdH}}
\begin{figure}[htb]
		\renewcommand{\thefigure}{S\arabic{figure}}
		\begin{center}
			\includegraphics[width = 0.8\linewidth ]{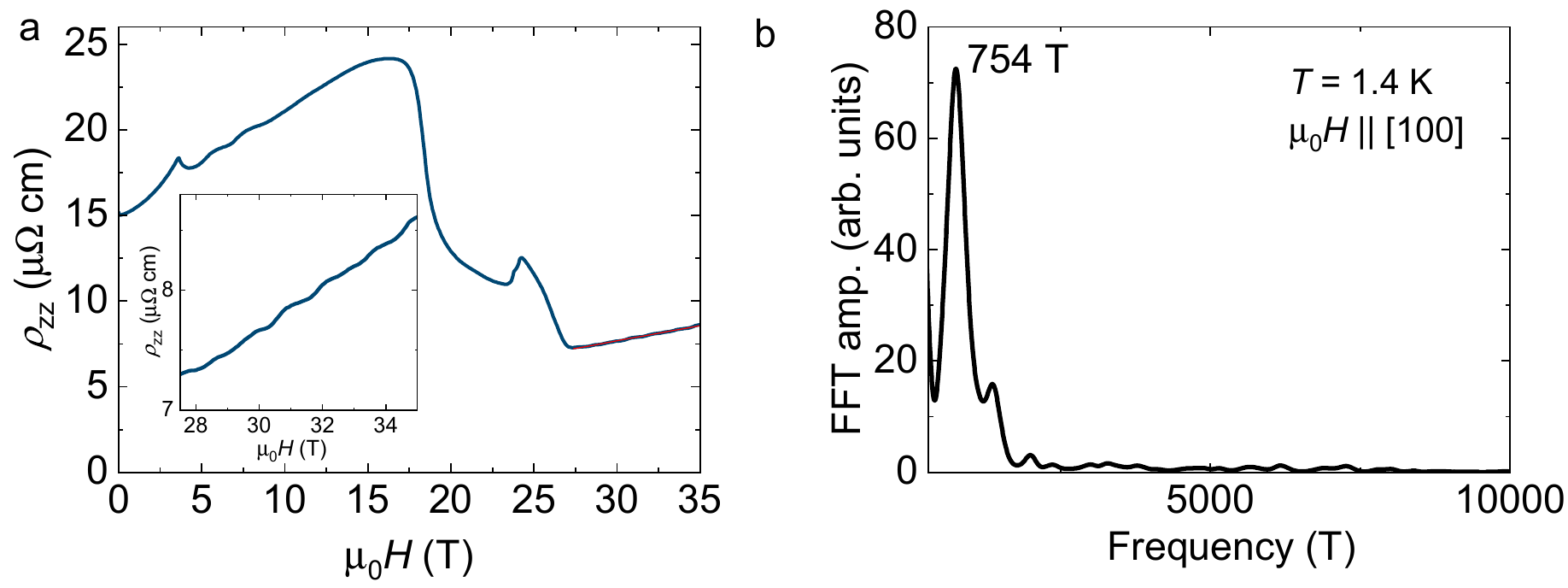}
			\caption{ (a) Magnetic field dependence of longitudinal resistivity at 1.4 K. The red line represents a linear fit in the FFM phase, where the oscillations are more pronounced. The inset provides a zoomed-in view of the high-field range. (b) Fast Fourier transform (FFT) of the oscillatory component of $\rho_{zz}$. } \label{QO}
		\end{center}
	\end{figure}

	LuMn$_6$Sn$_6$ showed clear Shubnikov-de Haas (SdH) oscillations 1.4~K. The oscillations are clearer in the FFM phase $\mu_0H>$27~T  aligned along the \textit{a}-axis. We analyzed the oscillations by subtracting a smooth polynomial background, and performing a fast Fourier transform (FFT) to get the component frequencies. The peak at 754~T in Fig.~\Ref{QO} corresponds to the extremal cross-sectional areas of the Fermi surface (FS). We used the Onsager relation $F=(\phi_0/2\pi^2)A_k$ to convert the FFT frequencies to reciprocal-space orbit cross section~\cite{shoenberg}. $\phi_0=2.07 \times 10^{-15}$ Tm$^2$ is the flux quantum, and $A_k$ is the cross-sectional area of the FS normal to the applied field.  A frequency of 754~T implies a FS pocket that encloses a $k$-space area (in the \textit{bc}-plane) of $A_k=7.2$~nm$^{-2}$. This is 10\% of the Brillouin zone of area $4\pi^2/bc=$76.9~nm$^{-2}$. 

\newpage

\renewcommand{\thesection}{S\arabic{section}}
	\section{Estimation of topological Hall resistivity \label{THE}}
	
    	\begin{figure*}[htb]
		\renewcommand{\thefigure}{S\arabic{figure}}
		\begin{center}
			\includegraphics[width = \linewidth ]{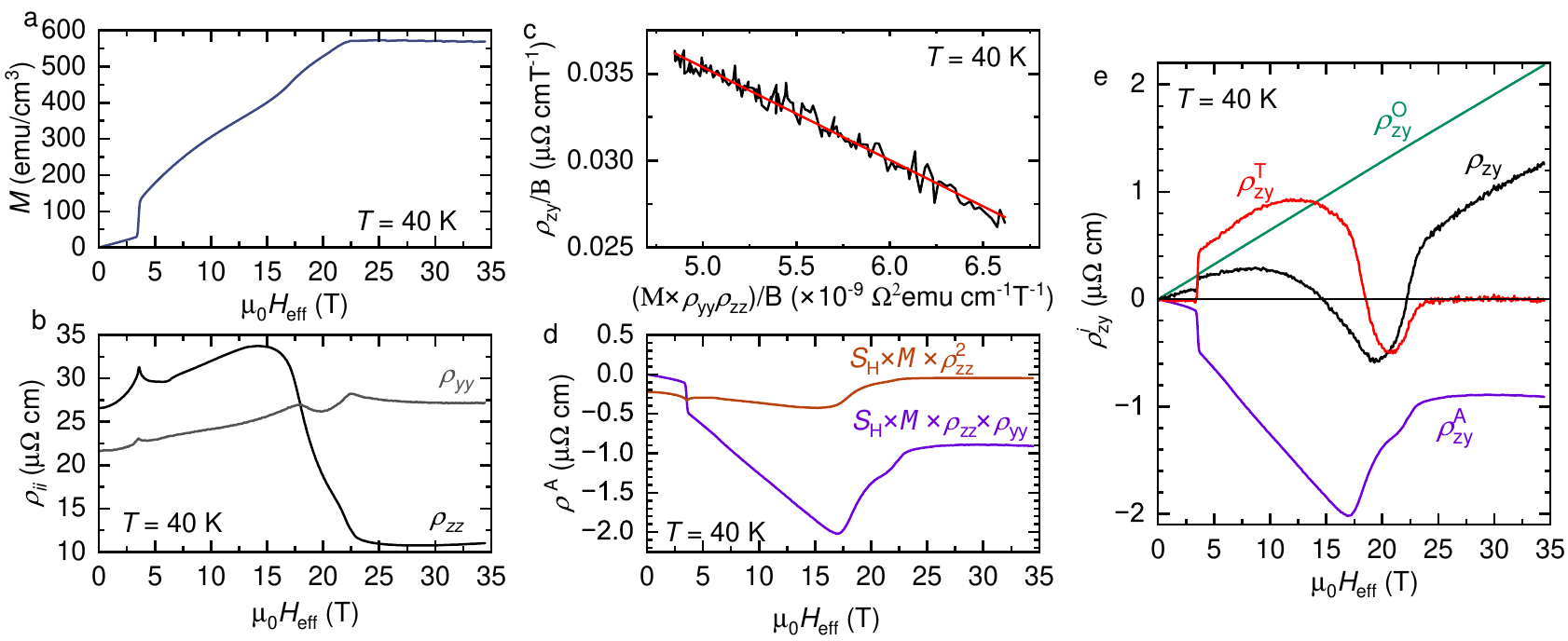}
			\caption{Estimation of topological Hall resistivity at 40~K for LuMn$_6$Sn$_6$. (a) and (b) show the field dependence of magnetization $M$, in-plane ($\rho_{yy}$) and out-of-plane ($\rho_{zz}$) longitudinal,  resistivity,   at $T=40$~K for $\textbf{H}\parallel a$. (c) $\frac{\rho_{zy}}{B}$ vs.\ $\frac{M}{B} \rho_{yy}\rho_{zz}$ curve at the forced ferromagnetic phase $\mu_0H > 25$ T. Red line is a linear fit to the experimental data. The intercept and slope correspond to $R_0$ and $S_H$, respectively. (d) Comparison of the anomalous Hall contribution obtained by $\rho^{A}=S_H M\rho_{zz}^2$ and $S_H M \rho_{yy}\rho_{zz}$, where the impact of anisotropy is considered. 
            (e) Estimated different components of $\rho_{zy}$. Curve labeled as $\rho_{zy}$ is the total Hall resistivity measured at 40~K. $\rho^{O}_{zy}$ and $\rho^{A}_{zy}$ are the normal Hall resistivity estimated from $R_0B$, and the anomalous Hall resistivity estimated from $S_H M \rho_{yy}\rho_{zz}$. $\rho^{T}_{zy}$ is the topological Hall resistivity derived from $\rho^{T}_{zy}=\rho_{zy}- \rho^{O}_{zy}-\rho^{A}_{zy}$.
            }

            \label{THE_explained}
		\end{center}.
	\end{figure*}
    
    For a noncoplanar magnetic metal, the empirical relation for the Hall effect resistivity is 
    
    \begin{equation}
        \rho_{zy} = \rho^{O}_{zy}+\rho^{A}_{zy}+\rho^{T}_{zy}=R_0B+ S_H M \rho_{yy}\rho_{zz} + \rho^{T}_{zy}
         \label{eq:rho_yz}
    \end{equation}

    The first term is the ordinary Hall effect ($\rho^{O}_{zy}=R_0B$) due to the Lorentz force. $R_0$ is ordinary Hall coefficient and $B=\mu_0(H_{\textrm{eff}}+M)$ is the induction field. $\mu_0H_{\textrm{eff}}=\mu_0(H-N_dM)$ is the effective magnetic field and $N_d$ is the demagnetizing factor of the sample and can be estimated from Ref.~\cite{demag_1,demag_2}. $\rho^{A}$ is the anomalous Hall resistivity, and $S_H$ is a constant. Here, we have used model~3 (see the main text) for the estimation of the AHE, as this model provides slightly better agreement with the expected absence of a THE in the AF1 phase (see main text Sec. IIIC2).
    The last term $\rho^{T}_{zy}$ represents the topological Hall resistivity. 

    At sufficiently high magnetic fields, in the forced ferromagnetic phase where the system is fully polarized, $\rho^{T}_{zy}$=0. $R_0$ and $S_H$ can be extracted in this regime by dividing both sides of Eq.~\ref{eq:rho_yz} by $B$, 

      \begin{equation}
        \frac{\rho_{zy}}{B} = R_0+ S_H \frac{M}{B} \rho_{yy}\rho_{zz}
         \label{eq:rho_yz_2}
    \end{equation}

    We plot the relationship between $\frac{\rho_{zy}}{B}$ vs.\ $\frac{M}{B} \rho_{yy}\rho_{zz}$ for scaling analysis. A representative curve at $T=40$~K is shown in Fig.~\Ref{THE_explained}(c). $R_0$ and $S_H$ were determined from the intercept and slope of the linear fit, respectively. 
    $\rho^{T}$ was estimated after the subtraction of $\rho^{O}_{zy}$ and $\rho^{A}$ from $\rho_{zy}$ as shown in Fig.~\Ref{THE_explained}(e).
    As pointed out in the main text, the field dependence of the in-plane ($\rho_{yy}$) and out-of-plane ($\rho_{yy}$) resistivity are distinct, as shown in Fig.~\ref{THE_explained}(b). Fig.~\ref{THE_explained}(d) further illustrates that if the anisotropy of magnetoresistivity is neglected and $\rho^{A}=S_H M\rho_{zz}^2$ is assumed, the resulting anomalous Hall effect will be much smaller, leading to an overestimation of the topological Hall effect. Moreover, we observe that employing the empirical relation $\rho^{A}=S_H M$  results in an even greater overestimation. 

    Similar procedure was used for the extraction of $\rho^{T}_{zy}$ at higher temperatures. Below 40~K, $\frac{\rho_{zy}}{B}$ vs.\ $\frac{M}{B} \rho_{yy}\rho_{zz}$ was no longer a linear line. Therefore, we were unable to extract $R_0$ and $S_H$.

\section*{References}
	
	\bibliography{Lu166}